\documentclass[prl,twocolumn,showpacs,superscriptaddress]{revtex4}
\usepackage{graphicx}
\usepackage{epsfig,color}
\usepackage{braket}
\usepackage{amsmath} 
\usepackage{amsthm} 
\usepackage{amssymb}    

\begin{document}
\newcommand{\ltwid}{\mathrel{\raise.3ex\hbox{$<$\kern-.75em\lower1ex\hbox{$\sim$}}}}
\newcommand{\gtwid}{\mathrel{\raise.3ex\hbox{$>$\kern-.75em\lower1ex\hbox{$\sim$}}}}
\newcommand{\sill}{\psi}
\newcommand{\trace}{{\rm Tr}}
\newcommand{\ntilde}{\tilde{n}}
\newcommand{\stilde}{\tilde{s}}
\newcommand{\atilde}{\tilde{\alpha}}
\newcommand{\greg}[1]{\textcolor{red}{#1}}
\newcommand{\adrian}[1]{\textcolor{blue}{#1}}
\newcommand{\george}[1]{\textcolor{cyan}{#1}}
\newcommand{\question}[1]{\textcolor{green}{#1}}

\title{Kondo versus indirect exchange: the role of the lattice and the actual range of RKKY interactions in real materials}

\author{Andrew Allerdt}
\affiliation{Department of Physics, Northeastern University, Boston, Massachusetts 02115, USA}
\author{C. A. B\"usser}
\affiliation{Department of Physics and Arnold Sommerfeld Center for Theoretical Physics, Ludwig-Maximilians-University Munich, Germany}
\author{G. B. Martins}
\affiliation{Department of Physics, Oakland University, Rochester, MI 48309, USA}
\author{A. E. Feiguin}
\affiliation{Department of Physics, Northeastern University, Boston, Massachusetts 02115, USA}

\date{\today}

\begin{abstract}
Magnetic impurities embedded in a metal interact via an effective Ruderman-Kittel-Kasuya-Yosida (RKKY) coupling mediated by the conduction electrons, which is commonly assumed to be long ranged, with an algebraic decay in the inter-impurity distance. However, they can also form a Kondo screened state that is oblivious to the presence of other impurities. The competition between these effects leads to a critical distance above which Kondo effect dominates, translating into a finite range for the RKKY interaction. We study this mechanism on the square and cubic lattices by introducing an exact mapping onto an effective one-dimensional problem that we can solve with the density matrix renormalization group method (DMRG). We show a clear departure from the conventional RKKY theory, that can be attributed to the dimensionality and different densities of states. In particular, for dimension $d>1$, Kondo physics dominates even at short distances, while the ferromagnetic RKKY state is energetically unfavorable. 
\end{abstract}

\maketitle
\paragraph{Introduction.} The Kondo problem describes a magnetic impurity screened by the spin of the electrons in the Fermi sea, 
forming a collective singlet state \cite{HewsonBook}. 
 This wave-funcion can be described as a hybridization cloud (``Kondo cloud'') centered at the impurity and decaying in distance with a characterisitic range $R_K$ \cite{Sorensen1996,affleck2009kondo,busser10}.
When more than one impurity interact with the conduction electrons, an effective Ruderman-Kittel-Kasuya-Yosida (RKKY) 
coupling between the magnetic moments arises \cite{RKKY1,RKKY2,RKKY3}, which can be ferro or antiferromagnetic, and oscillates with 
the distance between the impurities $R$ with wave-vector $2k_F$ (the Fermi momentum), and an amplitude that decays algebraically. 
It is commonly believed that if the Kondo screening length $R_K$ 
is shorter than the separation $R$, the Kondo effect will be 
more important and the RKKY interaction will not be observed. On the other hand, if $R$ is smaller than $R_K$, the RKKY interaction will dominate \cite{Sorensen1996,affleck2009kondo,busser10} .
As pointed out in Ref.~[\onlinecite{Sorensen1996}], even in a very dilute system with a low concentration of magnetic moments,
a finite number of impurities would be inside regions in space with overlapping Kondo clouds. The fact that Kondo physics dominates, and that a single impurity model can explain all experimental observations, clearly defies intuition.
The purpose of this work is to shed light on this issue by means of a numerical technique able to access the ground state of very large systems, and free of finite temperature effects. 

The Hamiltonian of the problem treated here is defined by two $S_i=1/2$ Kondo impurities (where $i=1,2$) interacting locally with free fermions in the bulk via an antiferromagnetic exchange coupling $J_K$:

\begin{equation}
H =  H_{\rm band} + J_K \left( \vec{S}_1 \cdot \vec{s}_{\rm r_1} + \vec{S}_2 \cdot \vec{s}_{\rm r_2} \right).
\label{hamiltonian}
\end{equation}
where $H_{\rm band}$ is the lattice Hamiltonian for non-interacting electrons, parametrized by a hopping $t$, and $\vec{s}_{r_i}$ represents the 
conduction electron's spin at the impurity's coordinate $r_i$, for impurities $i=1,2$.
As suggested by Doniach in Ref. \cite{Doniach1977} (see also \cite{Schwabe2012}), one could define a binding energy (or ``Kondo temperature'') for forming a Kondo singlet $T_K \simeq e^{-1/J_K}$, or an RKKY state, $T_{RKKY} \sim J_K^2$, and a competition between these two energy scales will dictate which phase will win.

The usual treatment to derive an effective exchange interaction between the localized moments involves second-order perturbation theory. The result can be summarized as:
\[
J_{RKKY}({\bf R})=J_K^2\chi({\bf R}),
\]
where $\chi({\bf R})$ is just the Fourier transform of the non-interacting static susceptibility, or Lindhard function. The dependence of this function on the distance varies with dimensionality. A universal expression is often offered in the literature, which is derived from assuming a uniform electron gas with a quadratic dispersion $E(k) \sim k^2$ \cite{Aristov1997}. Its asymptotic behavior at long distances ($k_F R \gg 1$) and in $d$ dimensions is of the form:
\[
\chi(R) \sim \frac{\sin{(2k_F R +\pi d/2)}}{R^d}.
\]
We note here that the effects of the lattice are completely ignored in this treatment. Clearly, the presence of a discrete lattice can have dramatic effects, due to the destructive and/or constructive interference of the electronic wave-functions centered on different sites, as well as the shape of both the dispersion and the Fermi surface \cite{RKKY3,Roth1966,Golosov1993,Schlottmann2000}. For instance, in graphene, the RKKY interaction can decay as $1/R^2$ for impurities sitting on lattice sites, or $1/R^3$ for impurities sitting at interstitial spaces \cite{brey2007,ando2005}.
%
%

\begin{figure}
  \centering
  \includegraphics[width=8.5cm]{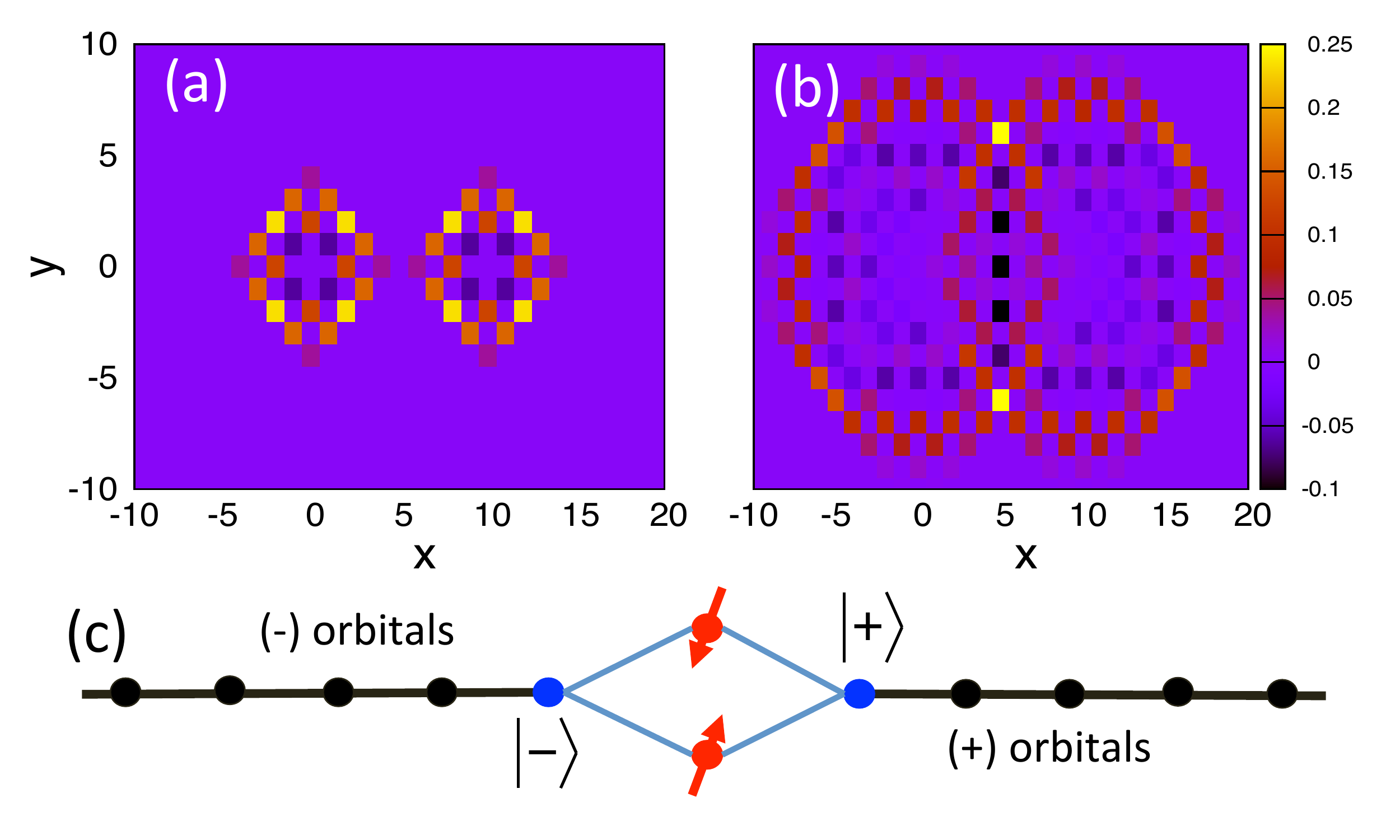}
  \caption{(color online) Examples of symmetric single-particle orbitals obtained through the Lanczos transformation for 
two impurities sitting at the origin, and at a distance $R=10$ along the $x$ direction, and after (a) 5 and (b) 10 iterations. 
In (c) we show the geometry of the equivalent problem, with the two magnetic impurities coupled to non-interacting 
tight-binding chains via many-body terms proportional to $J_K$ (more details in Supplemental Material \cite{supplemental}).
}
  \label{fig:orbitals}
\end{figure}

Irrespective of the dimensionality, a generic argument can show that on bipartite lattices and at half-filling, the oscillations in the RKKY interaction are commensurate with the lattice, and therefore, interactions are always ferromagnetic when moments are on the same sublattice, or antiferromagnetic otherwise \cite{saremi2007}. Therefore, the effects of perfect nesting, and the density of states(DOS) should be manifest in the strength of the interaction and its decay with distance, giving rise to a competition between Kondo and RKKY states that is non-universal and depends on the geometry and dimensionality of the system. 

\begin{figure}
  \centering
  \includegraphics[width=8.5cm]{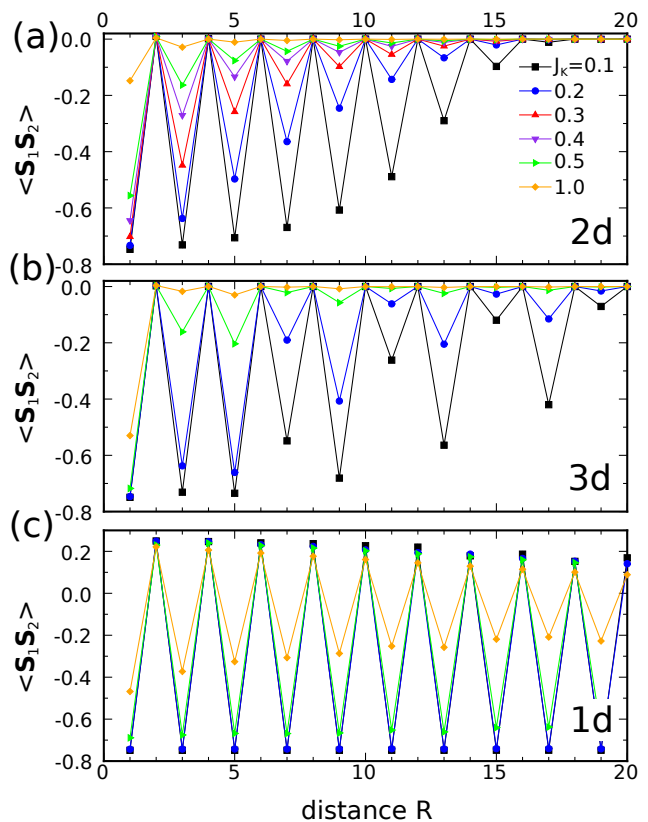}
  \caption{(color online) Spin-spin correlations between Kondo impurities at different distances along the $x$ axis, as 
a function of the Kondo coupling $J_K$ and for different lattice geometries: (a) 2d square, (b) 3d cubic and (c) 1d chain.
}
  \label{fig:corr}
\end{figure}

\paragraph{Method.} 
In this work we devise an approach to map the model in Eq.~(\ref{hamiltonian}) onto an effective one dimensional problem that is optimized for a DMRG calculation. 
By generalizing the method introduced in Ref.~\onlinecite{Busser2013} for single impurity problems, we reduce a complex lattice geometry to a single chain, or a multi-leg ladder in the case of multiple impurities.
A simple and straightforward analogy can be traced back to Wilson's numerical renormalization group (NRG) treatment of a single impurity coupled to a Fermi sea \cite{nrg_wilson,nrg_bulla}, where the electronic band is mapped onto a one-dimensional chain by means of a smart change of basis. 


We present two approaches to carry out the transformation. 
The first one relies on the so-called block Lanczos method \cite{cullum2002lanczos,qiao2005block}: 
We start the recursion by picking ``seed'' initial states, which are single-particle orbitals, from now on denoted $|1\rangle$, $|2\rangle$, 
centered at the position of the impurities, $\mathbf{r}_1$ and $\mathbf{r}_2$.
As shown in the Supplemental Material \cite{supplemental}, a block Lanczos method will generate a block tridiagonal 
matrix that can be interpreted as a single-particle Hamiltonian on a ladder geometry. 
The second approach applies to lattices with inversion symmetry: In this case we can simplify the problem even further by just defining two new seeds, which we take to be the symmetric and antisymmetric linear combinations of 
single-particle states $|\pm\rangle=1/\sqrt{2}\left(|1\rangle \pm |2\rangle \right)$. We then follow the prescription described in \cite{Busser2013} for the single 
impurity problem. By repeatedly applying on these states the non-interacting terms in the Hamiltonian, we generate new Lanczos orbitals in which the 
Hamiltonian has a tridiagonal form (see Fig. \ref{fig:orbitals}(a-b)). The two new sets of states generated by the two orthogonal seeds will also 
be orthogonal in this new basis, and the geometry of the problem is now reduced to two independent chains. As shown 
schematically in Fig.~\ref{fig:orbitals}(c), the magnetic impurities that originally are connected to orbitals $|1\rangle$ and $|2\rangle$, 
are now interacting with the $|\pm\rangle$ orbitals by complicated many-body terms that introduce a coupling between the two chains 
(For details we refer the reader to the Supplemental Material \cite{supplemental}). Nonetheless, the final Hamiltonian still is 
one dimensional and local, and its ground state can accurately be obtained using the DMRG method.
For the DMRG simulations, we take the total system size to be $L=4n$ (including impurities), 
since it has been already observed that Kondo does not develop in chains of length $L=4n+2$ \cite{Yanagisawa1991,hm09}. 
We have considered values of $L$ up to $204$, which corresponds approximately to a ``sphere'' around the two impurities of radius $\sim 100$, four times larger than the maximum inter-impurity distance considered in this work. As explained in Ref. [\onlinecite{Busser2013}], we assume ``infinite boundary conditions'', corresponding to orbitals that expand outward from the impurities and never hit any boundaries, which is equivalent to NRG, and a valid approach to study the thermodynamic limit. 
(Since we do not use a logarithmic discretization of the leads, we are still studying a ``Kondo box'' \cite{Thimm1999}. See Supplemental Material for details \cite{supplemental}).

\begin{figure}
  \centering
 \includegraphics[width=8.5cm,trim=0cm 0cm 0cm 5cm]{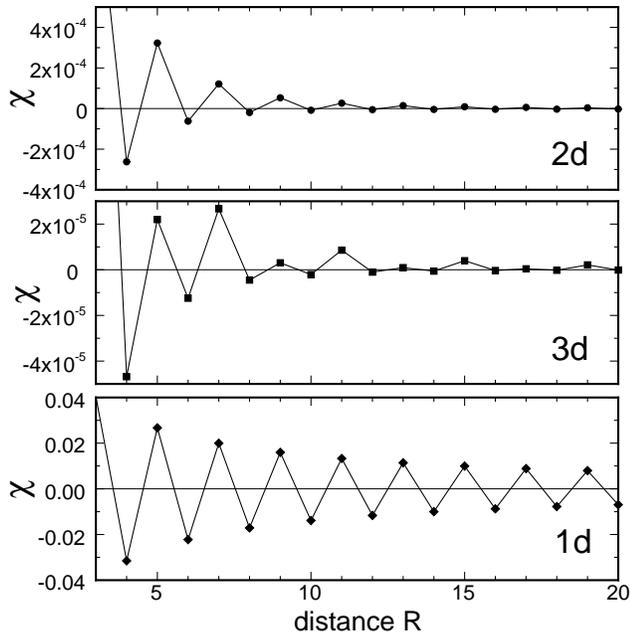}
  \caption{Lindhard function (spin susceptibility) for the non-interacting tight-binding model for the (from top to bottom) square, cubic lattices, and 1d chain.
Notice the different scales on the $y$ axes.
}
  \label{fig:chi}
\end{figure}

\paragraph{Results.} We have performed the mapping for two impurities embedded in square and cubic lattices, placing them at 
different distances $R$ along the horizontal $x$ axis. Unless otherwise specified, we typically show results for $L=124$. 
Fig.~\ref{fig:corr} shows the spin-spin correlations between both impurities as a 
function of $R$ for (a) square, (b) cubic lattices, and also (c) one-dimensional chain for comparison. In all three cases, we observe commensurate oscillations, and the 
different behaviors for impurities sitting at even or odd distances. First, we notice that ferromagnetic correlations at even distances are vanishingly small. This behavior has also been verified for impurities positioned along the diagonals of the lattice (not shown here). We focus our attention on the case of both impurities on different 
sublattices (at odd distances), and we find that for the 2d and 3d systems the correlations decay smoothly at first, but instead obeying an algebraic power-law, they 
have a marked change of behavior as they reach a crossover distance: for values of the interaction $J_K \simeq 0.1$, the impurities basically 
become uncorrelated for $R \simeq 20$ lattice spaces (or less, as $J_K$ increases). 


Results for the cubic lattice -- shown in Fig.~\ref{fig:corr}(b) -- display similar behavior as the square lattice, but 
with two important differences: the range of the correlations is slightly larger, and the amplitude of the oscillations has contributions from more 
than one mode, originating from the non trivial shape of the Fermi surface.
\cite{RKKY3,Roth1966,Golosov1993,Schlottmann2000}.
To see this explicitly, we just recall the expression for the Lindhard function
\begin{equation}
\chi(\mathbf{r}_1,  \mathbf{r}_2)=2\mathrm{Re}\sum\frac{\braket{\mathbf{r}_1 | n}\braket{n|\mathbf{r}_2}\braket{\mathbf{r}_2|m}\braket{m|\mathbf{r}_1}}{E_n-E_m},
\end{equation}
where the sum is over the eigenstates $n,m$ with energies $E_n>E_F>E_m$. 
The $\ket{\mathbf{i}}$ are the single-particle states at position $\mathbf{r}_i$, for $\mathbf{i}=1,~2$. 

We calculate this quantity numerically, and plot it for the square and cubic lattices in Fig.~\ref{fig:chi}, and we also include the 
one-dimensional case for comparison. We solved this formula explicitly with the mapping, and the exact eigenstates of 
a large system with both open and periodic boundary conditions, with indistinguishable results. The function displays the 
same oscillatory behavior as the spin-spin correlations. In particular, the ferromagnetic components for $R$ even are very 
weak compared to the antiferromagnetic counterpart.
We also notice a remarkable reduction by 2 and 3 
orders of magnitude in 2d and 3d, compared to the 1d case.

For this reason we now turn our attention to the one-dimensional case, which already has been studied in the literature \cite{Costamagna2008,Hallberg1997}. 
For consistency, we use our Lanczos transformation, keeping the total length of the system fixed at $L=124$. Results for the spin-spin correlations are 
shown in Fig.~\ref{fig:corr}(c), and are in sharp contrast to the higher dimensional examples: large values of $J_K$ are needed to induce a noticeable 
decay in the correlations. Moreover, for opposite sublattices, we obtain large positive values, indicating a ferromagnetic coupling, approaching the 
saturation value for small $J_K$. 

\begin{figure}
  \centering
  \includegraphics[width=8.5cm]{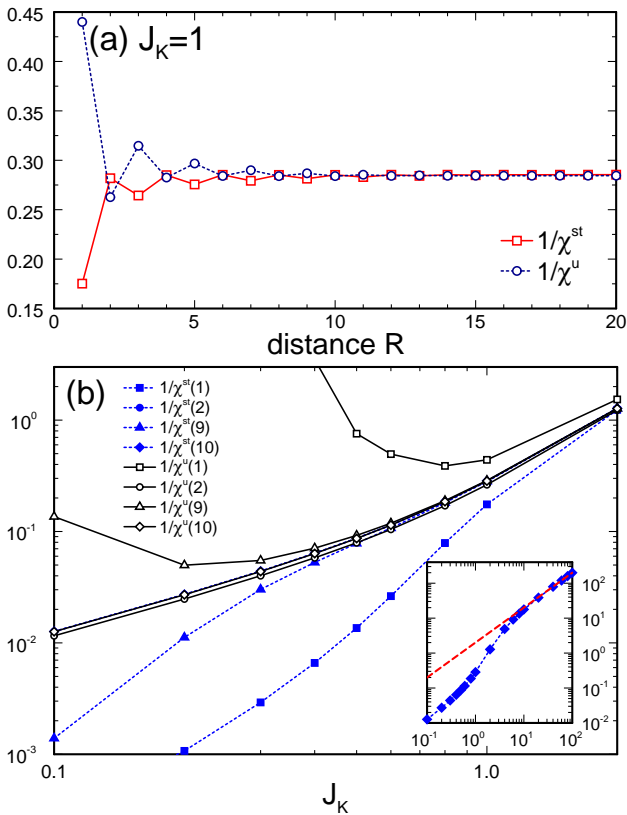}
  \caption{(color online) (a) Staggered and uniform impurity susceptibilities as a function of distance 
on the square lattice, for $J_K=1$. (b) Same quantities as a function of $J_K$, for distances $R=1,2,9,10$. The inset shows results for $R=10$ in an extended range, and the expression for a singlet for large $J_K$. Curves for $R=2,10$ are almost indistinguishable.
}
  \label{fig:binding}
\end{figure}

In order to determine whether the impurities are forming an RKKY singlet or not, we study the uniform and staggered magnetic susceptibilities for the 
impurities by numerically calculating $\chi ^{u,st}$ by applying a small 
(uniform or staggered) magnetic field of magnitude $h=10^{-4}$ to both impurities and evaluating $(d\langle S^z\rangle/dh)^{u,st}$ on one of them ($\langle \cdots \rangle$ means average taken over the ground state). In the universal Kondo regime, we expect $\chi^{u,st} \sim 1/T_K$. 
We show this quantity for a 2D square lattice in Fig.~\ref{fig:binding}(a). Results indicate that $\chi^{u}\simeq\chi^{st}$ for all even distances, and they asymptotically converge to the same 
value at long distances. 
This indicates that impurities on the same sublattice prefer to remain uncorrelated, in sharp contrast to the prediction that they would couple
ferromagnetically. Indeed, the FM state simply is energetically unfavorable.

We investigate this behavior as a function of $J_K$ in Fig.~\ref{fig:binding}(b). As seen for inter-impurity distance $R=9$, for 
instance, there is a crossover from an RKKY, to a Kondo regime at $J_K \simeq 0.5$. At distance $R=10$, the two susceptibilities are indistinguishable, while for $R=1,~2$, both clearly differ -- although slightly for $R=2$ --, signalling that impurities form a screened RKKY state.  
Asumming that $1/\chi^{u,st}$ define the binding energy scale $T_K, T_{RKKY}$, one would expect a crossover behavior -- from quadratic to exponential -- as a function of $J_K$, as suggested by the results for $R=9$. However, for $R=2,10$ we encounter that these quantities vary linearly as $\sim J_K$ for small $J_K$. This departure from the exponential form $T_K \sim \exp{(-1/J_K)}$ is in agreement with the analysis presented in Ref. \cite{Schwabe2012} and due to the discreteness of the spectrum, {\it i.e.} our system is a ``Kondo box'' with a level spacing of the order of $T_K$, and we are not in the universtal scaling regime \cite{Thimm1999,schlottmann2001kondo,simon2002finite,simon2003kondo,hand2006spin,Hanl2014}. For large $J_K$, as shown in the inset, the results asymptotically converge to the expression for a spin singlet $1/\chi = 2J_K$. This behavior deserves a detailed study, that will be presented elsewhere. 


\paragraph{Conclusions.} 
We studied the competition between RKKY and Kondo physics and the effect of dimensionality, by 
mapping the non-interacting Hamiltonian onto an effective one-dimensional lattice that can efficiently be 
solved using the DMRG method. We found a clear departure from the conventional picture: Above relatively short 
distances, a Kondo screened state becomes energetically favorable, and the impurities become completely uncorrelated. 
Moreover, the ferromagnetic state only develops in 1d, or weakly in higher dimensions and at very short distances. According to the behavior of the Lindhard function, and also the density-density correlation in the presence of a Kondo impurity \cite{Busser2013}, the probability of finding conduction electrons on two sites of the same sublattice can be vanishingly small and, as a consequence, their ability to mediate the RKKY interaction is greatly hindered.

This behavior is non-universal, and depends on the geometry of the lattice. In 3d, the RKKY correlations have a nontrivial oscillatory behavior due to the shape of the Fermi surface, that translates into contributions of several modes to the Lindhard function. Moreover, this function is one order of magnitude smaller in 3d than in 2d, but the range of the RKKY interactions is larger. This counterintuitive result may be an indication of non-perturbative effects.

Curiously, the density of states of a cubic lattice has a flat plateau spanning a range 
of energies $[-2,2]$ in units of the hopping $t$. This is identical to the flat DOS used in NRG calculations \cite{Jayaprakash1981,Jones1987}. 
However, for all ranges of couplings $J_K$ studied in 
this work, we have found important lattice effects, in agreement with previous quantum Monte Carlo (QMC) calculations \cite{Fye1989}. This illustrates the limitations of considering only spherical plane waves as the basis for constructing the NRG Hamiltonian \cite{nrg_bulla}. 

In particular, a remarkable result in early NRG studies of the two impurity problem \cite{Jones1988,Jones1989} -- with a linear dispersion and ignoring details of the lattice -- indicated the existence of a non-Fermi 
liquid critical point, characterized by a value of the spin correlations $\langle {\bf S}_1 \cdot {\bf S}_2 \rangle = -1/4$. 
Further QMC studies on two and three dimensional systems, with both a quadratic dispersion and a lattice, 
did not find any evidence of such a state \cite{Fye1987,Fye1989,Fye1994}. 
Later analysis revealed that the existence of such critical point required the presence of a very particular kind of particle-hole 
symmetry \cite{Affleck1992e,Affleck1995,Silva1996}, which is realized in our problem when $R$ is even. 
Our simulations have confirmed the QMC results, with a fast decay of the correlations, and the absence of anomalous behavior. 


Finally, we mention that our approach can readily be generalized to study realistic band structures, multi-orbital problems, 
and magnetic molecules, potentially bridging the gap between atomistic ab-initio calculations, and methods for strongly correlated problems.

\paragraph{Acknowledgments}
CAB was supported by the {\it Deutsche Forschungsgemeinschaft} (DFG) through FOR 912 under grant-no. HE5242/2-2 and through FOR 801.
GBM acknowledges financial support from the NSF under grants DMR-1107994 and MRI-0922811, and
AEF through grant DMR-1339564.

\bibliographystyle{apsrev}


\setcounter{figure}{0}
\makeatletter
\renewcommand{\thefigure}{S\@arabic\c@figure}

\section*{Supplemental material}
Here we describe the exact canonical transformation mapping the non-interacting Hamiltonian onto an equivalent geometry with reduced spatial dimensionality. In general, the total Hamiltonian of this problem is,
$$ H = H_\mathrm{band} + H_\mathrm{imp} + V $$
where $H_\mathrm{band}$ is the lattice Hamiltonian, $H_\mathrm{imp}$ is the many body impurity Hamiltonian ({\it e.g.}, 
Coulomb interactions in the case of Anderson impurities), and $V$ contains the hybridization terms coupling the lattice and the impurities. 
Here we give a general description of the method, without defining a particular lattice geometry. 
For clarity, we focus on the case of two magnetic impurities.

\subsection*{Block Lanczos method}
In order to generalize the Lanczos scheme proposed in Ref.~\onlinecite{Busser2013}, we propose two strategies, and will later show that they are intimately connected. 
The first technique consists of applying the extended block Lanczos method.

As done before for the single impurity, the first step is to choose the seed states. We will choose them to be single-particle orbitals 
sitting at the same lattice sites as the impurities, say sites $1$ and $2$. The advantage of this choice is that the hybridization terms in $V$ will 
remain unchanged under this transformation. The two initial states for the transformation are,
\begin{eqnarray}
\ket{\alpha_0}&=&c^{\dagger}_{1}\ket{0} \nonumber \\
\ket{\beta_0}&=&c^{\dagger}_{2}\ket{0}, \nonumber
\end{eqnarray}
where we have ignored the spin subindexes for simplicity.
A new set of states can be obtained using the extended Lanczos recursion method,

\begin{eqnarray}
\ket{\alpha_{n+1}} = H\ket{\alpha_n} &-&a^{\alpha\alpha}_n\ket{\alpha_n}-a^{\alpha\beta}_n\ket{\beta_n} \nonumber \\ 
&-&b^{\alpha\alpha}_n\ket{\alpha_{n-1}}-b^{\alpha\beta}_n\ket{\beta_{n-1}} \nonumber \\
\ket{\beta_{n+1}} = H\ket{\beta_n}&-&a^{\beta\beta}_n\ket{\beta_n}-a^{\beta\alpha}_n\ket{\alpha_n} \nonumber \\
&-&b^{\beta\beta}_n\ket{\beta_{n-1}}-b^{\beta\alpha}_n\ket{\alpha_{n-1}}. \nonumber
\end{eqnarray}

Requiring that the new states are orthogonal to the two previous states, {\it i.e.},
$$\braket{\alpha_{n-1}|\alpha_{n}}=0=\braket{\beta_{n-1}|\alpha_{n}}$$
results in the following equations that can be solved for the $b$ coefficients,

\begin{eqnarray}
\bra{\alpha_{n-1}}H\ket{\alpha_n}-b^{\alpha\alpha}_n\braket{\alpha_{n-1}|\alpha_{n-1}} - b^{\alpha\beta}_{n}\braket{\alpha_{n-1}|\beta_{n-1}} &=& 0 \nonumber \\
\bra{\beta_{n-1}}H\ket{\alpha_n}-b^{\alpha\alpha}_n\braket{\beta_{n-1}|\alpha_{n-1}}  - b^{\alpha\beta}_{n}\braket{\beta_{n-1}|\beta_{n-1}} &=& 0. \nonumber
\end{eqnarray}
A  similar set of equations determines the value of the $a$ coefficients,
\begin{eqnarray}
\bra{\alpha_{n}}H\ket{\alpha_{n}}-a^{\alpha\alpha}_n\braket{\alpha_{n}|\alpha_{n}} -a^{\alpha\beta}_{n}\braket{\alpha_{n}|\beta_{n}} & = & 0 \nonumber \\
\bra{\beta_{n}}H\ket{\alpha_n}-a^{\alpha\alpha}_n\braket{\beta_{n}|\alpha_{n}} -a^{\alpha\beta}_{n}\braket{\beta_{n}|\beta_{n}} & = & 0, \nonumber
\end{eqnarray} 
which reduces to 
$$\frac{\bra{\beta_{n}}H\ket{\alpha_{n}}}{\braket{\beta_{n}|\beta_{n}}} = a^{\alpha\beta}_{n},$$
because all $a^{\alpha\alpha}$ are zero.

So far, we have obtained a new non-normalized basis. However, states $\ket{\alpha_n}$ and $\ket{\beta_n}$ are not necessarily orthogonal. 
In order to obtain a full set of orthonormal states, we use a Gram-Schmidt procedure to orthogonalize these states (note that this is not the only choice).
\begin{eqnarray}
\ket{x_n} & = & \ket{\alpha_n} \nonumber \\
\ket{y_n} & = & \ket{\beta_n}-\braket{\alpha_n|\beta_n}\ket{\alpha_n}. \nonumber
\end{eqnarray}

Our Hamiltonian can now be written in the desired tridiagonal form:

$$H_\mathrm{band} =
 \begin{pmatrix}
  A_{0} & B_{1} & 0 & 0  & \cdots\\
  B_{1} & A_{1} & B_{2} & 0 & \\
  0  & B_{2}  & A_{2} & B_3 &  \\
  0 & 0 & B_3 & A_3 & \\ 
  \vdots & & & & \ddots
 \end{pmatrix},$$
where $A_n$ and $B_n$ are $2\times2$ matrices. This method can readily be extended further to more impurities. For $k$ impurities, each $A$ and $B$ matrix will be $k\times k$.
This matrix represents a new non-interacting tight-binding Hamiltonian, with each block representing a unit cell. For $k$ impurities, it can be recognized as $k$ coupled chains forming a $k\times L$ ladder. The new geometry is now quasi one-dimensional.

\begin{figure}
  \centering
  \includegraphics[width=8.5cm]{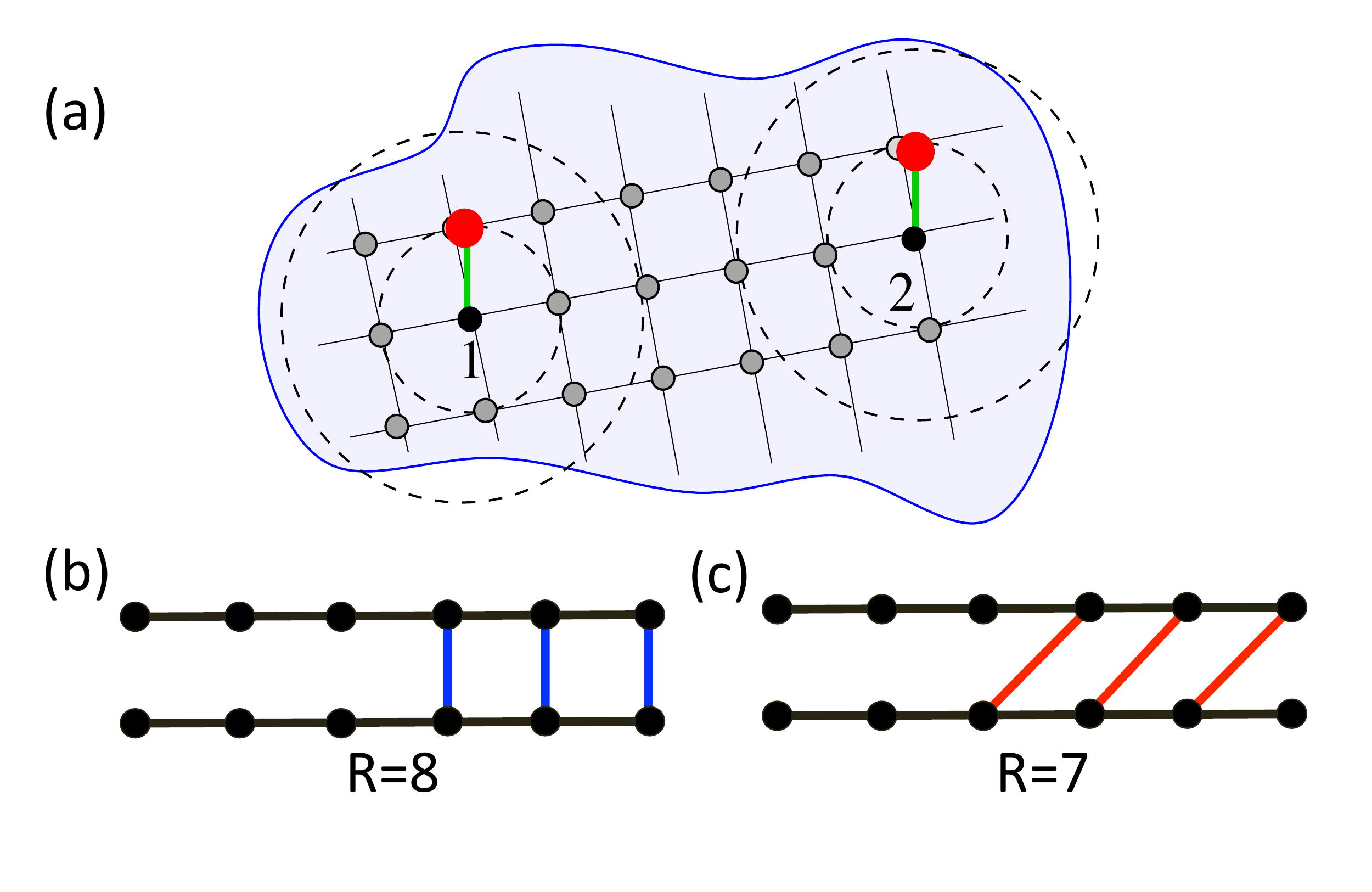}
  \caption{Geometry of the equivalent two impurity problem: (a) before and (b, c) after the transformation. (b) and (c) illustrate the cases of two impurities at distance $R=6$, and $R=5$, respectively. 
}
  \label{fig:ladder}
\end{figure}

We now turn to the particular case of the square lattice, and we look at what happens if the two impurities are on the same or opposite sublattices. In the first case, all the $A$ matrices will be zero, and the $B$ matrices will be lower triangular. This is represented in the slanted ladder shape of Fig. \ref{fig:ladder}(c).
On the other hand, if the impurities are on opposite sub-lattices, the $A$ matrices will be off-diagonal, while the $B$ matrices are diagonal. This is represented in Fig. \ref{fig:ladder}(b).

The first $R$ orbitals (where $R$ is the distance between the two impurities) will simply correspond to the single 
particle orbitals of two single-impurity problems, generating two independent chains. As the number of iterations 
increases, the orbitals will overlap, and the transformation will introduce mixing in order to preserve the orthogonality, leading to the 
hopping terms between both chains.

\subsection*{Bonding-antibonding symmetrization}

Now, lets turn our attention to a simple trick that will enable us to simplify the geometry of 
the equivalent problem even further in the case of lattices where we can define inversion symmetry. We just choose the initial states as linear combinations of single-particle orbitals
$$ \ket{\pm}=c^\dagger_\pm|0\rangle = \tfrac{1}{\sqrt{2}} (c^\dagger_1\pm c^\dagger_2) \ket{0},$$ 
representing symmetric (bonding) and antisymmetric (anti-bonding) states, respectively. For each initial state, 
the Lanczos iteration procedure is identical to that described in the single-impurity problem \cite{Busser2013}. Under 
this transformation, the many-body interactions in $V$ will be modified, introducing terms mixing the impurities, and the first two orbitals 
of both chains $\ket{\pm}$. However, the equivalent Hamiltonian will remain one-dimensional, and local, as shown in Fig.~1 in the main text.

After rotating all terms to the new basis, the many-body interactions acquire the form

\begin{eqnarray}
V &=& \frac{J_K}{2} \sum_{\lambda=\pm} \left( \mathbf{S}_1 + \mathbf{S}_2 \right) \cdot \sum_{\mu,\eta,\gamma=\pm} c^\dagger_{\gamma \mu} \vec{\sigma}_{\mu \eta} c_{\gamma \eta} \nonumber \\
 &+& \frac{J_K}{2} \sum_{\lambda=\pm} \left( \mathbf{S}_1 - \mathbf{S}_2 \right) \cdot \sum_{\mu,\eta,\gamma=\pm} c^\dagger_{\gamma \mu} \vec{\sigma}_{\mu \eta} c_{-\gamma \eta}.
\end{eqnarray}
Notice that this symmetrization is identical in spirit and form to the folding transformation used in NRG calculations for the two-impurity problem \cite{Jones1989}. The main 
difference is that our symmetrization takes place in real space, instead of momentum space.

\subsection*{Equivalence}

It is easy to show that both mappings are equivalent by simply taking any of the ladders 
in Fig.~\ref{fig:ladder}, and symmetrizing the orbitals under a reflection with respect to a plane parallel to 
the chains. The equivalent Hamiltonian will be nothing else but the chain in Fig.~1 in the text. As a consequence, 
the entanglement in the problem is reduced by a factor $2$, which translates into an exponential gain in terms of the number of states needed in the DMRG simulation. 

\subsection*{DMRG simulations}

In the block Lanczos approach, the bipartite entanglement is proportional to the number of legs, or impurities, in the problem. 
The main advantage of the folding transformation is that the entanglement gets reduced by two, and the number of states needed in the 
calculation is reduced by a power of $1/2$. For the DMRG simulations, we take a total system size to be $L=4n$ (including impurities), such that 
each impurity can form part of a collective RKKY state or its own Kondo cloud (it has been already observed that Kondo does not develop 
in chains of length $L=4n+2$ \cite{Yanagisawa1991}). We have considered values of $L$ up to $204$, which corresponds approximately to 
a ``sphere'' around the two impurities of radius $\sim 100$. As explained in Ref.~\onlinecite{Busser2013}, we assume ``infinite 
boundary conditions'', corresponding to orbitals that expand outward from the impurities and never hit any boundaries, which is valid if 
one is interested in the thermodynamic limit, in a similar spirit as the NRG approach. These system sizes are 4 times larger than the inter-impurity distance, and 
we have not observed significant finite size effects. Since the energy difference between the ground-state and the first excited state can 
be very small ($\sim 10^{-6}$), we fix the truncation error at $10^{-9}$ in all simulations, which translates into a number of 
DMRG states of the order of $3000$ or more in most cases.
Notice that this level of accuracy would be unattainable on the ladder geometry (without the bonding-antibonding symmetrization) due to the larger entanglement.


\end{document}